\documentclass[journal]{IEEEtran}
\ifCLASSINFOpdf
\else
\fi
\usepackage{amsmath,amssymb,enumerate,esint}
\usepackage{graphicx}
\usepackage{color}
\usepackage{placeins}
\usepackage{float}
\usepackage{hyperref}
\usepackage{tabularx,colortbl}
\newcommand{\abs}[1]{\left\vert{#1}\right\vert}
\usepackage{mathtools}
\DeclarePairedDelimiter{\ceil}{\lceil}{\rceil}

\begin{document}
\title{Input Admittance, Directivity and Quality Factor of Biconical Antenna of Arbitrary Cone Angle}

\author{Ramakrishna Janaswamy,~\IEEEmembership{Fellow,~IEEE}
        % <-this % stops a space
\thanks{R. Janaswamy is with the Department
of Electrical and Computer Engineering, University of Massachusetts, Amherst,
MA, 01003 USA e-mail: janaswam@umass.edu.}% <-this % stops a space
}


\markboth{IEEE Transactions on Antennas \& Propagation, submitted July 21, 2021, Revised November 09, 2021, Accepted November 11, 2021}%
{RJ\MakeLowercase{\textit{et al.}}: Input Admittance, Directivity and Quality Factor of Biconical Antenna of Arbitrary Cone Angle}



\maketitle

\begin{abstract}
New analytical expressions and numerical results for the mode coefficients, the directivity and the quality factor as well as computationally convenient expressions for the input admittance of a symmetrical biconical antenna of arbitrary length $L$ and cone angle $\theta_0$ are presented. The quality factor for a wide-angle biconical antenna is evaluated using three alternative formulations: (i) the evanescent energy stored outside the circumscribing sphere, (ii) the total evanescent energy stored in all space and (iii) by equivalent circuit model, and these are all compared with Chu's lower limit for an ideal antenna. Numerical calculations based on the analytical formula for antenna admittance confirm the conjecture that Foster's reactance theorem remains invalid even for perfectly conducting antennas. Furthermore, the variation of directivity of a wide-angle biconical antenna is a slowly varying function of its electrical length and is shown to depart significantly from that of a thin cylindrical dipole. Lastly, the ratio of directivity to $Q$ of an electrically small biconical antenna is shown to approach 78\% of the value of an ideal omnidirectional antenna. 
\end{abstract}


% \fi
\IEEEpeerreviewmaketitle

\section{Introduction}

\IEEEPARstart{T}{he} biconical antenna is one of the few finite-sized structures that is amenable to treatment by analytical methods. Furthermore, owing to its tapered shape, see Fig.~\ref{fg:Biconical},  it provides relatively wide bandwidths of operation as well as presents input impedance values that permit direct connection to practical feed lines. The antenna has previously been analyzed extensively by Schelkunoff \cite{Schelk}, \cite{Schelkunoff1952} and later by Tai \cite{Tai} and these works are summarized in \cite{CollZuck} and \cite{Jasik}. Even though these earlier works discuss the general field expressions around a biconical antenna, subsequent analysis and numerical data are provided only for small cone angles ($\theta_0\ll 1$) or for large cone angles ($\theta_0\to\pi/2$). The former special case is a good model for a thin cylindrical dipole and the latter to a spherical antenna with a thin equatorial slot. These two special cases not only reduce to elementary geometries but also permit analytical approximations that have been taken advantage of by these earlier workers and those that follow. Based on these approximations, input impedance and radiation patterns of wide-angle biconical antenna were studied in \cite{papas1} and \cite{papas2}. Bevensee \cite{Bevensee} generated extensive numerical data for input impedance and radiation patterns for cone angles up to $45$ degrees. Mayes \cite{Mayes} conducted experimental studies on tunable, wide-angle monopoles with loaded lumped elements to achieve specified bandwidth. Biconical antenna with unequal cone angles was treated in \cite{samaddar} and numerical results for antenna patterns were provided once again only for small and wide cone angles. 

Notwithstanding the above earlier works, given the practical importance of the antenna in a variety of wideband applications, including for calibration purposes \cite{sapuan} and its potential to serve as a benchmark antenna for computational methods \cite{janaswamy2021}, it is desirable to have systematic analysis steps leading to the generation of numerical data for various properties of the biconical antenna with arbitrary structural parameters. Furthermore, given that the canonical geometry of the biconical antenna permits an exact formulation of the electromagnetic boundary value problem, thus resulting in the availability of analytical expressions for the fields within and without the circumscribing sphere, the antenna offers a unique case for testing out the various forms of antenna $Q$-factors that have been advanced recently in the literature \cite{Yagh2005}, \cite{mclean}. This has not been explored previously. A second notion that can be assessed from the general expressions available for an antenna is the validity or not of the circuit-based Forster's reactance theorem, which has been ascertained in the past using only purely numerical methods \cite{Best2004}. Finally, exact field expressions permit investigation and application to a specific antenna geometry the generic limits governing the ratio of directivity to $Q$ \cite{chu1948}, \cite{gustafsson3}. The goal of the present paper is to provide this much-needed information and fill the void that currently exists in the literature. Non-availability of systematic analysis steps for the finite biconical antenna with arbitrary cone angle have prompted authors of modern antenna texts \cite{Stutzman}[p. 213], \cite{Balanis}[p. 500], \cite{jmjin} to treat it only peripherally, while dismissing the analysis as being too complicated. Here we provide these desirable steps, exact field expressions coupled with convenient numerical procedures for analyzing a general biconical antenna and generate numerical data for the various quantities for cone half-angle of $\theta_0 = \pi/4$. 

Of particular mention here are the following new contributions: (i) convergent expressions for the matrix entries involved in the determination of the mode spectrum (ii) convenient expressions for the determination of antenna terminal and input admittance from the mode spectra (iii) new expression for antenna directivity and (iv) detailed theory and new expressions involved in the determination of antenna quality factors, all valid for arbitrary cone angles and electrical lengths. In addition validated formulas are provided for the determination of eigenvalues (non-integer degrees of the Legendre functions) associated with the biconical structure that are needed in field expansions.

\section{Theory}
\subsection{Field Expressions and Basic Antenna Properties}

\begin{figure}[htb]
\centerline{\scalebox{0.45}{\includegraphics{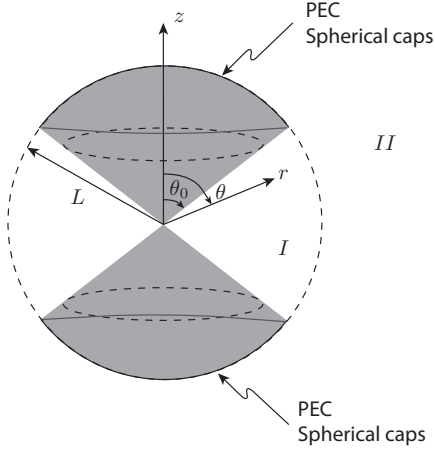}}}
\begin{center}
\caption{Symmetrical biconical antenna having radius $L$ and half-angle $\theta_0$. The cones are sealed off with perfectly conducting spherical caps.}
\label{fg:Biconical}
\end{center}
\end{figure}

A symmetric, coaxial biconical antenna having identical, perfectly conducting (PEC) cones with half angles $\theta_0$ and axes aligned is shown in Fig.~\ref{fg:Biconical}. The region between the two cones is defined by $\theta_0\le \theta\le \pi-\theta_0$, $0<r\le L$. The cones are terminated in spherical PEC caps at the two ends. An infinitesimal source is assumed to be placed at the apex of the antenna. 

We assume operation in a balanced mode where the feed electric field near the apex is an even function of polar angle $\theta$ about $\theta = \pi/2$. The only non-zero components of field in such a case are $E_r,E_\theta,$ and $H_\phi$ and these are all azimuthally invariant. Such a field constitutes low order TM$_r$ modes, that is, those that have no azimuthal variation. The input admittance, $Y_{\rm in}$, of the antenna is defined as the ratio of apex current $I(0)=:I_0$ to the voltage difference between the two cones $V(0)=:V_0$ 
\begin{equation}
Y_{\rm in} = \frac{I_0}{V_0} := \lim\limits_{r\to 0}\frac{2\pi r\sin\theta H_\phi}{\int\limits_{\theta_0}^{\pi-\theta_0} rE_\theta\,d\theta}.\label{eq:Yip}\end{equation}
For TM$_r$ modes (non-TEM modes), the complete fields can be obtained in terms of the radial electric field, $E_r$, by an application of Bouwkamp-Casimir theorem \cite[p. 1-29]{RJ2020}. 

For an $e^{j\omega t}$ time convention at the angular frequency $\omega$, the fields in region-I ({\em viz.}, for $r<L$) are expressed as
\begin{eqnarray}
rE_r&=&-\frac{j\eta I_0}{2\pi kr}\sum\limits_{\nu_n}a_{\nu_n}\frac{\widehat{J}_{\nu_n}(kr)}{\widehat{J}_{\nu_n}(kL)}M_{\nu_n}(\theta)\label{eq:ErI}\\
rE_\theta&=&-\frac{j\eta I_0}{2\pi}\sum\limits_{\nu_n}\frac{a_{\nu_n}}{\nu_n(\nu_n+1)}\frac{\widehat{J}_{\nu_n}'(kr)}{\widehat{J}_{\nu_n}(kL)}\frac{\partial M_{\nu_n}(\theta)}{\partial\theta}\nonumber\\
&&+\ rE_0(r,\theta)\label{eq:EthI}\\
rH_\phi&=&-\frac{I_0}{2\pi}\sum\limits_{\nu_n}\frac{a_{\nu_n}}{\nu_n(\nu_n+1)}\frac{\widehat{J}_{\nu_n}(kr)}{\widehat{J}_{\nu_n}(kL)}\frac{\partial M_{\nu_n}(\theta)}{\partial\theta}\nonumber\\
&&+\ rH_0(r,\theta)\label{eq:HphI},
\end{eqnarray}
where $\widehat{J}_{\nu_n}(\cdot)$ is the spherical Riccati-Bessel function of order, $\nu_n$, $M_{\nu_n}(\theta) = [P_{\nu_n}(\cos\theta)-P_{\nu_n}(-\cos\theta)]/2$ is an odd function of $\theta$ about $\theta=\pi/2$, $P_\nu(\cdot)$ is the Legendre function of degree $\nu$, $\eta = \sqrt{\mu_0/\epsilon_0}$ is the intrinsic impedance and $k = \omega\sqrt{\mu_0\epsilon_0}$ is the wavenumber in free-space at the operating frequency $\omega$. A prime on Bessel functions denotes derivative with respect to its entire argument. 

The fields $E_0, H_0$ correspond to outgoing and incoming radial TEM modes and are assumed to be of the form
\begin{eqnarray}
rE_0(r,\theta) &=&\frac{I_0\eta}{4\pi\sin\theta}\bigg[\left(1+KY_t\right)e^{-jk(r-L)}\nonumber\\
&&+\left(1-KY_t\right)e^{jk(r-L)}\bigg]\label{eq:TEMe}\\
rH_0(r,\theta)&=&\frac{I_0}{4\pi\sin\theta}\bigg[\left(1+KY_t\right)e^{-jk(r-L)}\nonumber\\
&&-\left(1-KY_t\right)e^{jk(r-L)}\bigg]\label{eq:TEMh}.
\end{eqnarray}
The constant $K$ is the characteristic impedance of an infinite biconical transmission line and is related to the half-angle of the cone via
\begin{equation}
K = \frac{\eta}{2\pi}\int\limits_{\theta_0}^{\pi-\theta_0}\frac{1}{\sin\theta}\,d\theta = \frac{\eta}{\pi}\ln\left[\cot\left(\frac{\theta_0}{2}\right)\right],\label{eq:chimp}
\end{equation}
and the constant $Y_t$ is the terminal admittance at $r=L$ such that 
\begin{equation}
KY_t = \frac{\eta H_0(L,\theta)}{E_0(L,\theta)}.\label{eq:KYt}
\end{equation}
From the field expressions of $E_0$ and $H_0$ it is clear that $2\pi r\sin\theta E_0(L,\theta) = I_0\eta$ and $2\pi r\sin\theta$ $H_0(L,\theta) = I_0KY_t$ so that  (\ref{eq:KYt}) follows directly from these. Note that the non-TEM parts of the electric and magnetic fields in region-I do not contribute to the voltage difference and current at the apex that appear in the definition of input admittance (\ref{eq:Yip}). This is due to the small argument behavior of the Bessel function $\widehat{J}_\nu(z)\sim O(z^{\nu+1})$ and the fact that $\nu>0$ here. Consequently, the numerator in (\ref{eq:Yip}) is independent of $\theta$ as it should.

In order that $E_r = 0$ on the conical surfaces at $\theta = \theta_0, \theta = \pi-\theta_0$, the order $\nu_n$ must be a root of $M_{\nu_n}(\theta_0) = 0$, that is, of 
\begin{equation}
P_{\nu_n}(\cos\theta_0) - P_{\nu_n}(-\cos\theta_0) = 0,\ n = 1,2,\ldots.\label{eq:roots1}
\end{equation}
Clearly, the distribution of roots depends only on the half-angle $\theta_0$ of the biconical antenna and independent of its length. Equation (\ref{eq:roots1}) will have a countably infinite number of roots $\nu_n>0$ as can be easily ascertained from the asymptotic form of Legendre function \cite[p. 419]{Schelk2} subject to $q = \sqrt{\left(\nu_n+\frac{1}{2}\right)^2+\frac{1}{4}}\gg 1$

\begin{eqnarray}
M_{\nu_n}(\theta) &\sim& \sqrt{\frac{2}{\pi\sin\theta\sqrt{\nu_n(\nu_n+1)}}}\sin\left(\frac{q\pi}{2}-\frac{\pi}{4}\right)\nonumber\\
&&\sin\left[q\left(\frac{\pi}{2}-\theta\right)\right].\label{eq:Masymp}\end{eqnarray}
Using this asymptotic form, the non-integer zeros of the function $M_{\nu_n}(\theta_0)$ and the corresponding derivatives are easily determined as 
\begin{eqnarray}
\nu_n&\sim&\sqrt{n^2\alpha_0^2-\frac{1}{4}}-0.5\sim n\alpha_0-0.5\label{eq:Rootsasymp1}\\
\frac{d\nu_n}{d\theta_0}&\sim&\frac{n^2\alpha_0^3}{\pi(\nu_n+\frac{1}{2})}\sim\frac{n\alpha_0^2}{\pi},\ n=1,2,\ldots\label{eq:Rootsasymp2}
\end{eqnarray}
where $\alpha_0 = \pi/(\pi/2-\theta_0)$. For a cone half-angle of $\theta_0 = \pi/4$, the slopes of linear trends in equations (\ref{eq:Rootsasymp1}) and (\ref{eq:Rootsasymp2}) for $\nu_n$ and $d\nu_n/d\theta_0$ are respectively, $\alpha_0=4$ and $\alpha_0^2/\pi \approx 5.0930$. 

The Legendre function $M_{\nu_n}(\theta)$ can also be expressed in terms of a hypergeometric function $F(\cdot)$ and the Gamma function $\Gamma(\cdot)$ as \cite[p. 178]{Lebedev}
\begin{eqnarray}
M_{\nu_n}(\theta) &=& \frac{2\Gamma\left(\frac{\nu_n}{2}+1\right)}{\sqrt{\pi}\Gamma\left(\frac{\nu_n}{2}+\frac{1}{2}\right)}\sin\left(\frac{\nu_n\pi}{2}\right)\ \cos\theta\nonumber\\
&&\times\ F\left(\frac{1}{2}-\frac{\nu_n}{2},\frac{\nu_n}{2}+1;\frac{3}{2};\cos^2\theta\right),
\end{eqnarray}
so that $\nu_n$ is also a root of 
\begin{equation}
F\left(\frac{1}{2}-\frac{\nu_n}{2},\frac{\nu_n}{2}+1;\frac{3}{2};\cos^2\theta_0\right)=0,\ n = 1,2,\ldots.\label{eq:Roots}
\end{equation}
The roots $\nu_n$ of the equation (\ref{eq:Roots}) for a given value of $\theta_0$ can be calculated numerically by generating the hypergeometric function $F(\cdot)$ from its governing differential equation \cite[Equation 9.151]{GrRy2007}. The derivative $d\nu_n/d\theta_0$ can also be calculated numerically using the difference formula $d\nu_n/d\theta_0 \approx [\nu_n(\theta_0+\delta\theta_0)-\nu_n(\theta_0)]/\delta\theta_0$. 

\begin{figure}[htb]
\centerline{\scalebox{0.45}{\includegraphics{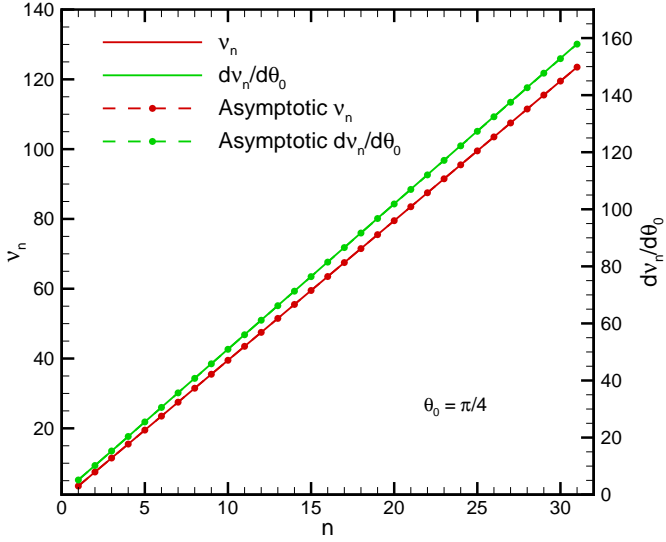}}}
\begin{center}
\caption{First thirty one roots of equation (\ref{eq:Roots}) and zeros of (\ref{eq:Masymp}).}
\label{fg:Roots}
\end{center}
\end{figure}

The formulas given in (\ref{eq:Rootsasymp1}), (\ref{eq:Rootsasymp2}) for the roots and the derivatives are very accurate over a wide range of cone angles. To demonstrate this we show in Table~\ref{Tb:Roots} the first thirty roots and the associated derivatives for $\theta_0 = \pi/4$ using $\delta\theta_0 = 5\times 10^{-4}\,$rad. These numerically generated values from equation (\ref{eq:Roots}) are compared with the asymptotic values given in (\ref{eq:Rootsasymp1}) and (\ref{eq:Rootsasymp2}). Figure~\ref{fg:Roots} shows a plot of the roots $\nu_n$ and the derivative $d\nu_n/d\theta_0$ versus $n$ for the numerically and asymptotically generated values. The discrete data are joined by a smooth curve for visual clarity. Excellent agreement is seen between the two sets of data. A linear trend is clearly seen for both $\nu_n$ and $d\nu_n/d\theta_0$ with $n$. For a given cone angle ($\theta_0 = \pi/4$ here), the numerically calculated values of $\nu_n$ and $d\nu_n/d\theta_0$ can be fit by linear regression to result in 
\begin{equation}
\nu_n = 4.0005n-0.5135;\ \frac{d\nu_n}{d\theta_0}=5.0954n+2.8012\times 10^{-2}.\label{eq:Rootsfit}
\end{equation}
Asymptotic formula based values for the slope and intercept for the linear trends in $\nu_n$ and $d\nu_n/d\theta_0$ given after equation (\ref{eq:Rootsasymp2}) are in excellent agreement with the numerically generated ones in (\ref{eq:Rootsfit}) and the relative error for the roots and its derivatives are less than $1.25\times 10^{-2}\%$ and $4.8\times 10^{-2}\%$, respectively. Clearly, the asymptotic values of the roots are already very accurate (less than 0.2\,\% error) even for $n=1$ and get increasingly better for higher $n$. The asymptotic formulas (\ref{eq:Rootsasymp1}) and (\ref{eq:Rootsasymp2}) are also effective for very low cone angles. For instance, for the extremely small value of $\theta_0 = 1^o\approx 0.0175\,$radians, the actual root as computed numerically is $\nu_1=1.26$, whereas that predicted by formula (\ref{eq:Rootsasymp1}) is $\nu_1\sim 1.46$. Accuracy will only increase as the root order increases. It is noteworthy that the coefficients in formulas (\ref{eq:Rootsasymp1}) and (\ref{eq:Rootsasymp2}) are dependent on the cone-angle and that the functional dependence is attested here by favorable comparison with the numerically generated ones for the special case of $\theta_0=\pi/4$. In summary formulas (\ref{eq:Rootsasymp1}) and (\ref{eq:Rootsasymp2}) can be safely employed over a wide range of cone angles.
\begin{table}[h]
\caption{\bf First thirty roots of equation (\ref{eq:Roots}) for $\theta_0 = \frac{\pi}{4}$.}
\centering
{\scriptsize \begin{tabular}{ccccc}
\hline\hline
Index $n$&Numerical $\nu_n$&Numerical $\frac{d\nu_n}{d\theta_0}$&Asymptotic $\nu_n$&Asymptotic $\frac{d\nu_n}{d\theta_0}$\\[0.25ex]
\hline
1&3.4620574e+00&5.2057059e+00&3.4686270e+00&5.1332193e+00\\
2&7.4804257e+00&1.0220051e+01&7.4843597e+00&1.0205869e+01\\
3&1.1486850e+01&1.5285263e+01&1.1489579e+01&1.5292155e+01\\
4&1.5490107e+01&2.0368186e+01&1.5492186e+01&2.0381787e+01\\
5&1.9492073e+01&2.5490328e+01&1.9493749e+01&2.5472752e+01\\
6&2.3493388e+01&3.0597680e+01&2.3494791e+01&3.0564383e+01\\
7&2.7494329e+01&3.5698304e+01&2.7495535e+01&3.5656393e+01\\
8&3.1495036e+01&4.0795498e+01&3.1496094e+01&4.0748640e+01\\
9&3.5495586e+01&4.5890810e+01&3.5496528e+01&4.5841045e+01\\
10&3.9496027e+01&5.0985049e+01&3.9496875e+01&5.0933561e+01\\
11&4.3496387e+01&5.6078677e+01&4.3497159e+01&5.6026157e+01\\
12&4.7496688e+01&6.1171979e+01&4.7497396e+01&6.1118814e+01\\
13&5.1496942e+01&6.6265145e+01&5.1497596e+01&6.6211517e+01\\
14&5.5497160e+01&7.1358308e+01&5.5497768e+01&7.1304257e+01\\
15&5.9497349e+01&7.6451570e+01&5.9497917e+01&7.6397025e+01\\
16&6.3497515e+01&8.1545013e+01&6.3498047e+01&8.1489818e+01\\
17&6.7497661e+01&8.6638706e+01&6.7498162e+01&8.6582630e+01\\
18&7.1497791e+01&9.1737112e+01&7.1498264e+01&9.1675458e+01\\
19&7.5497907e+01&9.6836166e+01&7.5498355e+01&9.6768300e+01\\
20&7.9498012e+01&1.0193433e+02&7.9498437e+01&1.0186115e+02\\
21&8.3498106e+01&1.0703175e+02&8.3498512e+01&1.0695402e+02\\
22&8.7498192e+01&1.1212855e+02&8.7498580e+01&1.1204689e+02\\
23&9.1498271e+01&1.1722483e+02&9.1498641e+01&1.1713977e+02\\
24&9.5498343e+01&1.2232069e+02&9.5498698e+01&1.2223265e+02\\
25&9.9498409e+01&1.2741621e+02&9.9498750e+01&1.2732555e+02\\
26&1.0349847e+02&1.3251148e+02&1.0349880e+02&1.3241844e+02\\
27&1.0749853e+02&1.3760657e+02&1.0749884e+02&1.3751134e+02\\
28&1.1149858e+02&1.4270155e+02&1.1149888e+02&1.4260425e+02\\
29&1.1549863e+02&1.4779648e+02&1.1549892e+02&1.4769716e+02\\
30&1.1949867e+02&1.5289142e+02&1.1949896e+02&1.5279007e+02\\
\hline
\label{Tb:Roots}
\end{tabular}}
\end{table}

The remaining unknown coefficients $a_{\nu_n}$ in (\ref{eq:ErI})-(\ref{eq:HphI}) are determined by an application of the boundary conditions at $r=L$, which will be elaborated upon below. The input admittance as defined in (\ref{eq:Yip}) is 
\begin{equation}
Y_{\rm in}=\lim_{r\to 0}\frac{2\pi r\sin\theta H_\phi}{\int\limits_{\theta_0}^{\pi-\theta_0}rE_\theta\,d\theta}=\frac{1}{K}\frac{1-\Gamma_{\rm in}}{1+\Gamma_{\rm in}},\label{eq:BiconInAdm}\end{equation}
where $\Gamma_{\rm in}$ is the reflection coefficient at $r=0$ of the TEM mode. It is obtained from the total field expressions in (\ref{eq:EthI}), (\ref{eq:TEMe}), (\ref{eq:HphI}) and (\ref{eq:TEMh}) as 
\begin{equation}
\Gamma_{\rm in} = \frac{1-KY_t}{1+KY_t}e^{-2jkL}=:\Gamma_L e^{-j2kL},\label{eq:Gamin}\end{equation} 
where $\Gamma_L$ is the reflection coefficient of the TEM mode at the truncating surface $r = L$.

The fields in region-II ({\em viz.}, for $r>L$) are expressed as 
\begin{eqnarray}
rE_r&=&\frac{-j\eta I_0}{2\pi kr}\sum\limits_{\ell=1,3}^\infty
\ell(\ell+1)c_\ell \frac{\widehat{H}^{(2)}_\ell(kr)}{\widehat{H}^{(2)}_\ell(kL)}P_\ell(\cos\theta)\label{eq:ErII}\\
rE_\theta&=&-\frac{jI_0\eta}{2\pi}\sum\limits_{\ell=1,3}^\infty c_{\ell}\frac{\widehat{H}_{\ell}^{(2)\prime}(kr)}{\widehat{H}_{\ell}^{(2)}(kL)}\frac{d P_{\ell}(\cos\theta)}{d\theta}\label{eq:EthII}\\
rH_\phi&=&-\frac{I_0}{2\pi}\sum\limits_{\ell = 1,3}^\infty c_{\ell}\frac{\widehat{H}_{\ell}^{(2)}(kr)}{\widehat{H}_{\ell}^{(2)}(kL)}\frac{d P_{\ell}(\cos\theta)}{d\theta}\label{eq:HphII}
\end{eqnarray}
where $\widehat{H}_{\ell}^{(2)}(\cdot)$ is the spherical Riccati-Hankel function of the second  kind of order $\ell$, $P_\ell(\cos\theta)$ is a Legendre function of first kind of integer degree $\ell$ and $c_\ell$ are the mode constants to be determined by an enforcement of boundary conditions at $r=L$. As with the interior region, the fields $E_\theta$ and $H_\phi$ are both even functions of $\theta$ about $\theta = \pi/2$, while $E_r$ is an odd function. Bouwkamp and Casimir's theorem \cite[p. 1-29]{RJ2020} guarantees that there will be no TEM waves in region-II\footnote{This is because non-trivial fields {\em outside} a circumscribing sphere require either a non-zero $E_r$ or a non-zero $H_r$.}, so the expansions given in (\ref{eq:ErII})-(\ref{eq:HphII}) are complete for region-II. The individual terms contained within the summation signs of (\ref{eq:ErII})-(\ref{eq:HphII}) constitute the TM$_{\ell,0}$ modes. 

The far-zone fields are obtained by setting 
$\widehat{H}_\ell^{(2)'}(kr)\sim j^{\ell+1}e^{-jkr},\ kr\gg 1$, resulting in
\begin{equation}
rE_\theta=\eta rH_\phi\sim\frac{I_0\eta e^{-jkr}}{2\pi}\sum\limits_{\ell=1,3}^\infty \frac{c_{\ell}j^\ell}{\widehat{H}_{\ell}^{(2)}(kL)}\frac{d P_{\ell}(\cos\theta)}{d\theta}.\label{eq:Biconfar}
\end{equation}

By matching the tangential fields $E_\theta$ and $H_\phi$ at $r=L$ the following relations between the expansion coefficients ${c}_\ell$ and $a_{\nu_n}$ are obtained
\begin{equation}
\sum\limits_{\ell = 1,3}^\infty c_\ell\frac{d P_\ell(\cos\theta)}{d\theta}=\begin{cases}
\sum\limits_{n=1}^\infty\frac{a_{\nu_n}\frac{\partial M_{\nu_n}}{\partial\theta}}{\nu_n(\nu_n+1)} -\frac{KY_t}{\sin\theta},&{\scriptstyle\theta_0<\theta<\pi-\theta_0}\\
0,&\rm otherwise\end{cases}\label{eq:cofrel1}
\end{equation}
and 
\begin{eqnarray}
\sum\limits_{\ell = 1,3}^\infty c_\ell\frac{\widehat{H}_\ell^{(2)\prime}(kL)}{\widehat{H}_\ell^{(2)}(kL)}\frac{d P_\ell(\cos\theta)}{d\theta}=\qquad\qquad&&\nonumber\\
\begin{cases}\displaystyle
\sum\limits_{n=1}^\infty\frac{a_{\nu_n}\frac{\widehat{J}_{\nu_n}^\prime(kL)}{\widehat{J}_{\nu_n}(kL)}\frac{\partial M_{\nu_n}}{\partial\theta}}{\nu_n(\nu_n+1)} +\frac{j}{\sin\theta},&{\scriptstyle\theta_0<\theta<\pi-\theta_0}\\
0,&\rm otherwise\end{cases}&&\label{eq:cofrel2}
\end{eqnarray}

Orthogonal properties between the various kernel functions inside the summation sign may now be exploited to get an explicit relation between the coefficients $c_\ell$ and $a_{\nu_n}$. The following integral relations are easy to establish based on the governing differential equation for Legendre function \cite[p. D-15]{RJ2020}:
\begin{equation}
L_{\ell,n} =\int\limits_0^\pi P_\ell(\cos\theta)P_n(\cos\theta)\sin\theta\,d\theta
=\frac{2}{2n+1}\delta_\ell^n\label{eq:PlPn}\end{equation}
\begin{equation}
L^\prime_{\ell,n}=\int\limits_0^\pi\frac{dP_\ell(\cos\theta)}{d\theta}\frac{dP_n(\cos\theta)}{d\theta}\sin\theta\,d\theta= \frac{2n(n+1)}{2n+1}\delta_\ell^n\label{eq:dPldPn}\end{equation}
\begin{eqnarray}
T_{\ell,\nu_n}(\theta_0) &=&\frac{1}{2}\int\limits_{\theta_0}^{\pi-\theta_0}P_\ell(\cos\theta)M_{\nu_n}(\theta)\sin\theta\,d\theta \nonumber\\
&=&\frac{\sin\theta_0P_\ell(\cos\theta_0)\displaystyle\left.\frac{\partial M_{\nu_n}}{\partial\theta}\right\rvert_{\theta_0}}{\nu_n(\nu_n+1)-\ell(\ell+1)}\label{q:Tln}\end{eqnarray}
\begin{eqnarray}
I_{\nu_n,\nu_m}(\theta_0)&=&\frac{1}{2}\int\limits_{\theta_0}^{\pi-\theta_0}\frac{\partial M_{\nu_m}(\theta)}{\partial\theta}\frac{\partial M_{\nu_n}(\theta)}{\partial\theta}\sin\theta\,d\theta\label{eq:Inumu1}\\
&=&\frac{\nu_n(\nu_n+1)}{2}\int\limits_{\theta_0}^{\pi-\theta_0}M_{\nu_m}(\theta)M_{\nu_n}(\theta)\sin\theta\,d\theta\label{eq:Inumu2}\\
&=&\delta_n^m\frac{\nu_n(\nu_n+1)}{(2\nu_n+1)}\sin\theta_0\left.\left[\frac{\partial M_{\nu_n}}{\partial\theta}\right]^2\left[{\frac{d\nu_n}{d\theta}}\right]^{-1}\right\rvert_{\theta_0},\nonumber
\end{eqnarray}
where $\delta_n^m$ is the Kronecker's delta equal to 1 if $m=n$ and zero otherwise. Making use of these relations in appropriate integrals with respect to $\theta$ of (\ref{eq:cofrel1}) and (\ref{eq:cofrel2}) and carrying out several algebraic manipulations we are led to  
\begin{equation}
 KY_t = \frac{1}{\ln\left[\cot\left(\frac{\theta_0}{2}\right)\right]}\sum\limits_{\ell = 1,3}^\infty c_\ell P_\ell(\cos\theta_0), \label{eq:BiconAdm}
\end{equation}
\begin{equation}
a_{\nu_n} = \frac{\nu_n(\nu_n+1)}{I_{\nu_n,\nu_n}(\theta_0)}\sum\limits_{\ell =1,3}^\infty \ell(\ell+1)c_\ell T_{\ell,\nu_n}(\theta_0),\label{eq:atob}
\end{equation}
and 
\begin{eqnarray}
&&\sum\limits_{\ell = 1,3}^\infty\Biggl\{\Big[m(m+1)\ell(\ell+1)\sin\theta_0P_m(\cos\theta_0)  P_\ell(\cos\theta_0)g_{m\ell}\nonumber\\
&&-\frac{m(m+1)}{(2m+1)}\frac{\widehat{H}_m^{(2)\prime}(kL)}{\widehat{H}_m^{(2)}(kL)}\delta_m^\ell\Big]c_\ell\Biggr\}
= jP_m(\cos\theta_0),\label{eq:cmeq}
\end{eqnarray}
where 
\begin{eqnarray}
g_{m\ell} &=& g_{\ell m}=\sum\limits_{n=1}^\infty\Biggl\{\frac{(2\nu_n+1)}{\nu_n(\nu_n+1)}\frac{\widehat{J}^\prime_{\nu_n}(kL)}{\widehat{J}_{\nu_n}(kL)}\frac{d\nu_n}{d\theta_0}\times \label{eq:gml1}\\
&&
\frac{1}{[\nu_n(\nu_n+1)-m(m+1)][\nu_n(\nu_n+1)-\ell(\ell+1)]}\Biggr\}.\nonumber
\end{eqnarray}
Equation (\ref{eq:cmeq}) describes an infinite linear system for the determination of the coefficients $c_\ell$ given the electrical length $kL$ and the angle $\theta_0$ of the biconical antenna. 

To speed up computation of the quantities $g_{ml}$ in the series (\ref{eq:gml1}), it is possible to extract out the asymptotic tail of the series and rewrite it is 
\begin{eqnarray}
g_{m\ell} &=& \sum\limits_{n=1}^\infty\Bigg\{\frac{(2\nu_n+1)}{\nu_n(\nu_n+1)}\frac{\widehat{J}_{\nu_n}'(kL)}{\widehat{J}_{\nu_n}(kL)}\frac{d\nu_n}{d\theta_0}\nonumber\\
&&\frac{1}{[\nu_n(\nu_n+1)-\alpha_0^2\beta_m^2][\nu_n(\nu_n+1)-\alpha_0^2\beta_\ell^2]}\nonumber\\
&&-\frac{1}{\pi\alpha_0^2(\beta_m+\beta_\ell)kL}\biggl[\frac{1}{(n-\beta_m)(n-\beta_\ell)}-\nonumber\\
&&\frac{1}{(n+\beta_m)(n+\beta_\ell)}\biggr]\Bigg\}\nonumber\\
&&+\frac{1}{\pi\alpha_0^2(\beta_m+\beta_\ell)kL}\biggl[\frac{\pi\sin\pi(\beta_m-\beta_\ell)}{(\beta_m-\beta_\ell)\sin\pi\beta_m\sin\pi\beta_\ell}+\nonumber\\
&&\frac{1}{\beta_m\beta_\ell}-2\frac{\psi(\beta_m)-\psi(\beta_\ell)}{\beta_m-\beta_\ell}\biggr],\hskip2.5em
\label{eq:Newgml}\end{eqnarray}
where $\beta_m=\sqrt{m(m+1)}/\alpha_0$ and $\psi(x)=d\ln\Gamma(x)/dx$ is the Euler psi-function (a.k.a. digamma function) \cite[8.360-1]{GrRy2007}. Note that (\ref{eq:Newgml}) is still an exact expression but the series in (\ref{eq:Newgml}) converges much faster than that in (\ref{eq:gml1}), thereby permitting accurate and rapid computation of the associated mode coefficients $c_\ell$. As an illustration of the convergence properties, the terms within the series in (\ref{eq:Newgml}) behave as $O(n^{-3.8}),\  n\gg 1$ for $m = 5,\ell = 9$, $\theta_0 = \pi/4$, and $kL = \pi$. 

The infinite linear system of equations contained in (\ref{eq:cmeq}) can be cast in a matrix form as \begin{equation}{\cal G}{\bf c} = {\bf p},\label{eq:BCA_Coeff}\end{equation} where the vector ${\bf c} = [c_1,c_3,c_5,\ldots]^T,$ the vector 
\begin{equation}
{\bf p} = j[P_1(\cos\theta_0), P_3(\cos\theta_0), P_5(\cos\theta_0),\ldots]^T\label{eq:vecp}\end{equation}
and the symmetric matrix ${\cal G} = \{{\cal G}_{m\ell}\}$ has entries 
\begin{eqnarray} 
{\cal G}_{m\ell} &=&  m(m+1)\ell(\ell+1)\sin\theta_0P_m(\cos\theta_0)P_\ell(\cos\theta_0)g_{m\ell}\nonumber\\
&&-\frac{m(m+1)}{2m+1}\frac{\widehat{H}_m^{(2)\prime}(kL)}{\widehat{H}_m^{(2)}(kL)}\delta^\ell_m.\label{eq:CalG}\end{eqnarray}
For a given angle $\theta_0$ and electrical length $kL$, the coefficients $c_\ell$ can be determined by solving a truncated version of the linear system (\ref{eq:BCA_Coeff}) with entries given in (\ref{eq:CalG}), (\ref{eq:Newgml}) and (\ref{eq:vecp}). The coefficients $a_{\nu_n}$ are then determined from these and (\ref{eq:atob}). All components of fields in the entire space are completely determined once these mode spectra are known. All of these calculations are made possible for an arbitrary cone angle by  the availability of formulas (\ref{eq:Rootsasymp1}) and (\ref{eq:Rootsasymp2}) for the eigenvalues.

\begin{figure}[htb]
\centerline{\scalebox{0.45}{\includegraphics{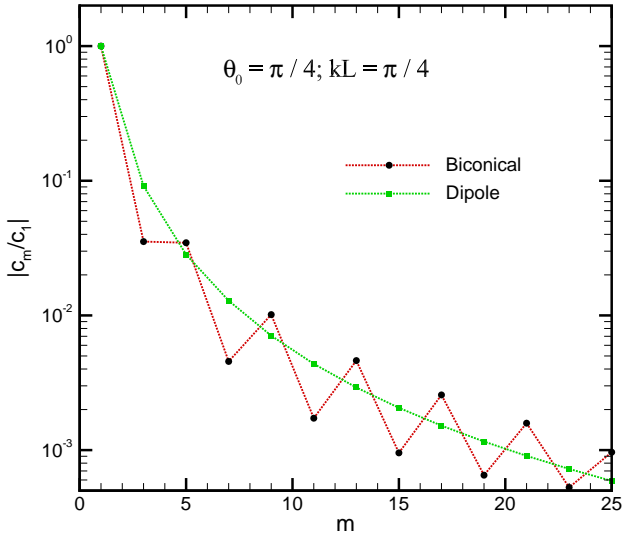}}}
\begin{center}
\caption{The normalized mode coefficients $\abs{c_m/c_1}, \abs{c_m^d/c_1^d}$,
$m = 1,3,\ldots$ for the biconical antenna and the short dipole antenna.}
\label{fg:BCA_Coeff}
\end{center}
\end{figure}

It is interesting to compare the radial electric field of a biconical antenna with that of a center-fed, linear electric dipole of length $2L$ placed with its center at the origin and oriented along the $z$-axis. Assuming a triangular current distribution $I(z) = I_0(1-\abs{z}/L)$ (valid for a short electric dipole) and using the spherical harmonic expansion outside the circumscribing sphere \cite[p. 1-25]{RJ2020} and the small argument approximation of the spherical Bessel function $j_\ell(x)\sim \frac{2^{\ell}\ell!}{(2\ell+1)!}x^\ell$ that is valid with an error less than $4\,\%$ for $x\le \pi/5$, the radial electric field, $E_r^d$, in the region $r>L$ of the short electric dipole can be obtained as 
\begin{equation}
rE_r^d = -\frac{\eta I_0}{kr}\sum\limits_{\ell = 1,3,\ldots}^\infty \sqrt{\frac{2\ell+1}{\pi}}\frac{(2kL)^\ell\ell\,!}{(2\ell+1)!}\widehat{H}_\ell^{(2)}(kr)P_\ell(\cos\theta).\label{eq:Erd}
\end{equation}
Comparing with the expression (\ref{eq:ErII}) we define the corresponding mode coefficient, $c_\ell^d$, for the short electric dipole as 
\begin{equation}
c_\ell^d = \sqrt{\frac{2\ell+1}{\pi}}\frac{\widehat{H}_\ell^{(2)}(kL)}{\ell(\ell+1)}\frac{(2kL)^\ell\ell\,!}{(2\ell+1)!}.\label{eq:celldip}
\end{equation}
Fig.~\ref{fg:BCA_Coeff} shows the normalized values of mode coefficients $\abs{c_m/c_1}$ calculated by truncating the linear system (\ref{eq:BCA_Coeff}) to the first 13 odd modes (that is, $m\in(1,3,\ldots,25)$) for $kL = \pi/4$ and $\theta_0 = \pi/4$. Clearly the coefficient strengths beyond $m=11$ diminish below 1\% of the dominant mode. Also shown in the plot is the normalized mode spectrum $\abs{c_m^d/c_1^d}$ for the short electric dipole. It is seen that the short biconical antenna has roughly the same mode content as that of a short dipole {\em outside the circumscribing sphere}.

The real and imaginary parts of the mode coefficients for a longer antenna with $kL = \pi$ are shown in Table~\ref{Tb:Coeff} as an illustration. For this relatively longer antenna, the mode coefficient $c_{39}$ is seen to be four orders of magnitude less than the dominant one. In general highly accurate results were found by considering $M_i \sim  10\ceil{kL}$ exterior modes so that the size of the truncated matrix $\cal G$ is of the order of $M_i\times M_i$. The infinite series in (\ref{eq:Newgml}) can also be truncated to $N \sim 10\ceil{kL}$. 
\begin{table}[h]
\caption{\bf Mode coefficients $c_m$ for $\theta_0 = \frac{\pi}{4}, kL = \pi$.}
\centering
\begin{tabular}{ccc}
\hline\hline
Coefficient Index $m$&Real ($c_m$)& Imag ($c_m$)\\[0.25ex]
\hline
1&1.1482413e+00&8.0494496e-02\\
3&-3.3991143e-02&-2.0680854e-01\\
5&-4.0128576e-02&-1.1550678e-01\\
7&-9.3402703e-03&4.2614872e-03\\
9&8.9429512e-03&2.5641567e-02\\
11&2.8255266e-03&-4.9111323e-04\\
13&-4.0929315e-03&-1.0946386e-02\\
15&-1.3274186e-03&-1.0991429e-04\\
17&2.3225998e-03&5.9470173e-03\\
19&7.7230276e-04&2.4380244e-04\\
 21&-1.4816312e-03&-3.6858693e-03\\
 23&-5.1340186e-04&-2.7350591e-04\\
 25&1.0179272e-03&2.4832825e-03\\
 27&3.7578307e-04&2.7866791e-04\\
 29&-7.3539008e-04&-1.7715199e-03\\
 31&-2.9829691e-04&-2.8413689e-04\\
 33&5.4947727e-04&1.3158048e-03\\
 35&2.5895960e-04&3.0774407e-04\\
 37&-4.1701878e-04&-1.0028417e-03\\
39&-2.7858839e-04&-4.3547987e-04\\

\hline
\label{Tb:Coeff}
\end{tabular}
\end{table}

\begin{figure}[htb]
\centerline{\scalebox{0.45}{\includegraphics{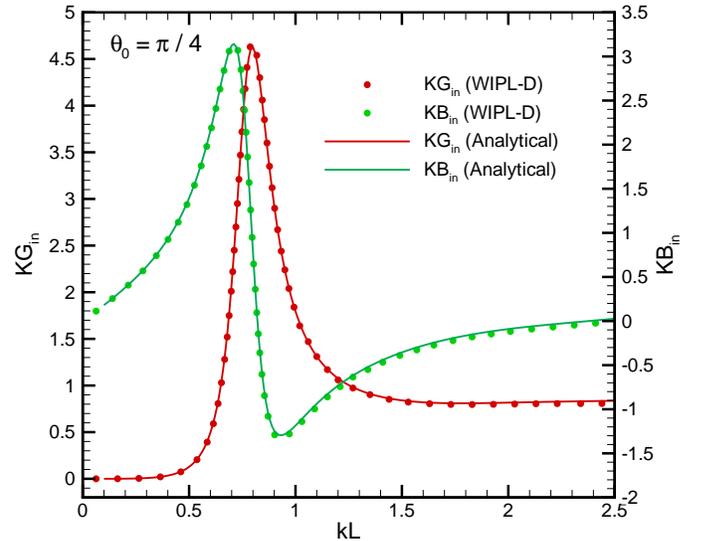}}}
\begin{center}
\caption{Comparison of the normalized input admittance $KY_{\rm in} = K(G_{\rm in}+jB_{\rm in})$ of the biconical antenna for $\theta_0 = \pi/4$. The constant $K = 105.765\,\Omega$.}
\label{fg:BCA_Admin}
\end{center}
\end{figure}

The real and imaginary parts of the normalized input admittance of the antenna $KY_{\rm in}=K(G_{\rm in}+jB_{\rm in})$ as calculated analytically from (\ref{eq:BiconInAdm}), (\ref{eq:Gamin}) and (\ref{eq:BiconAdm}) are  plotted in Fig.~\ref{fg:BCA_Admin} as a function of $kL$ for $\theta_0=\pi/4$. Also shown for comparison are results from the commercial software package WIPL-D\footnote{A delta-gap generator defined on a cylindrical wire of radius $\Delta/60$, where $\Delta$ was the size of the gap in the central part of the antenna was used to feed the structure. A second feed model was also chosen with a point generator defined at the point where wire radius was equal to zero. In the point gap model, the wire itself had conical shape, with one radius equal to $0$ and the other radius equal to the radius of cone at the point where wire connects to the plate part of the model. Results for the input admittance with these two models were reported to be practically identical \cite{Branko}.}. Excellent agreement is seen between the two calculations. For this value of cone angle, the input admittance is seen to vary only slowly beyond $kL = 1.25$, thus exhibiting wide-band behavior for sufficiently long antenna lengths. 

From complex Poynting theorem applied to the entire space, we know that if $P_s = \frac{1}{2}V_0I_0^*$ is the complex power supplied by the source and if the time-averaged electric and magnetic energies stored in the entire volume $V$ are, respectively, $W_e=\frac{\epsilon_0}{4}\int_V{\abs{\bf E}}^2\,dv$  and $W_m=\frac{\epsilon_0}{4}\int_V{\abs{\eta{\bf H}}}^2\,dv$ then
\begin{equation}
-\Im[{P_s}] = \frac{{\abs{I_0}}^2B_{\rm in}}{2{\abs{Y_{\rm in}}}^2} = 2\omega(W_e-W_m),\label{eq:Pcons}
\end{equation}
where the operator $\Im(\cdot)$ stands for the `imaginary part of' a quantity. At resonance $B_{\rm in} = 0$ and $W_e=W_m$. The biconical antenna in Fig.~\ref{fg:BCA_Admin} exhibits first resonance at $kL \approx 0.8$, below which it is capacitive (that is, stored electric energy in all space exceeds stored magnetic energy), while it is seen to be inductive beyond the resonance through to $kL \approx 2.5$. These facts have important bearing on the determination of the $Q$-factor in the ensuing analysis.

It is also seen that the frequency derivative of the susceptance, $dB_{\rm in}/dk$, is not a monotonous function of frequency. Fig.~\ref{fg:BCA_Admin} shows that the frequency derivative is positive for $0.1< kL<0.7$ and $0.9<kL<2.5$. However, it turns negative for $0.7\le kL\le 0.9$. These results confirm the conjecture \cite{Capek} that Foster's reactance theorem \cite[Ch. 8]{RJ2020}, which was developed for lossless passive networks does not hold even for perfectly conducting antennas. Such an invalidity of the Foster's reactance theorem to antennas was also established analytically in \cite[Sec. 8.1.3]{RJ2020} for the case of an equatorial slot cut on the surface of a perfectly conducting sphere.  

\begin{figure}[htb]
\centerline{\scalebox{0.4}{\includegraphics{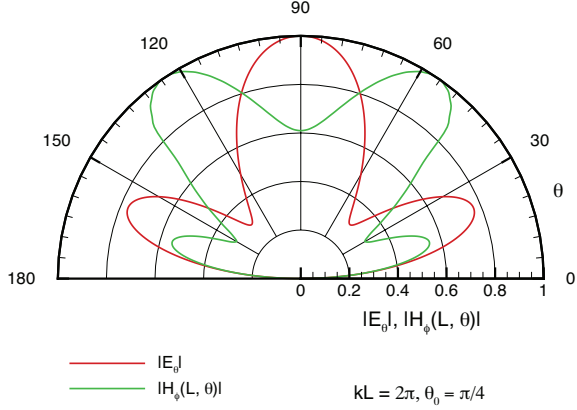}}}
\begin{center}
\caption{Far-zone $|E_\theta(\theta)|$ and near-zone $|H_\phi(r=L,\theta)|$ of the biconical antenna for $\theta_0 = \pi/4, kL = 2\pi$.}
\label{fg:BCA_Pattern}
\end{center}
\end{figure}

The E-plane variation of far-zone electric field $E_\theta$ and near-zone magnetic field $H_\phi(r=L,\theta)$ of the antenna with $kL = 2\pi, \theta_0 = \pi/4$ are shown in Fig.~\ref{fg:BCA_Pattern}. Both fields are normalized with respect to their maximum fields. The far-zone pattern for an electrically small biconical antenna resembles that of short dipole. However, for the pattern shown in Fig.~\ref{fg:BCA_Pattern} for a radius of $L/\lambda = 1$, a single-lobed broadside beam is observed for the biconical antenna. This is in contrast to a two wavelength long cylindrical dipole, which will have split beams \cite{Stutzman}.
\subsection{Radiated Power and Directivity}
Using the field expressions (\ref{eq:ErII})-(\ref{eq:HphII}) and the identities (\ref{eq:PlPn}), (\ref{eq:dPldPn}), the complex power flowing out of a sphere of radius $r > L$ can be shown to be 
\begin{eqnarray}
P_{\rm cf}(r) &=&\frac{1}{2}\oiint\limits_\Omega r^2E_\theta H_\phi^*\,d\Omega\label{eq:Pcflow}\\
&=&j\abs{I_0}^2\eta\sum\limits_{\ell=1,3}^\infty\frac{\ell(\ell+1)\abs{c_\ell}^2}{2\pi(2\ell+1)}\frac{\widehat{H}^{(2)\prime}_\ell(kr)\widehat{H}_\ell^{(1)}(kr)}{\abs{\widehat{H}_\ell^{(2)}(kL)}^2},\nonumber
\end{eqnarray}
where $\Omega$ denotes the unit sphere. The real part of the complex power flow gives the power radiated, $P_{\rm rad}^+$, in the  volume exterior to the antenna. The superscript $^+$ denotes that fields exterior to $r=L$ have been used in obtaining it.  The normalized radiated power is 
\begin{equation}
P_o^+=\frac{2\pi P_{\rm rad}^+}{\eta\abs{I_0}^2}=\sum\limits_{\ell = 1,3}^\infty\frac{\ell(\ell+1)}{(2\ell+1)}\frac{\abs{c_\ell}^2}{\abs{\widehat{H}^{(2)}_\ell(kL)}^2},\label{eq:PoII}
\end{equation}
where the Wronskian for the spherical Riccati-Hankel function $W_H = \widehat{H}_\ell^{(2)\prime}(z) \widehat{H}^{(1)}_\ell(z)-\widehat{H}_\ell^{(1)\prime}(z)\widehat{H}_\ell^{(2)}(z) = -2j$ has been used in arriving at the result.  The radiated power is seen to be independent of the radius $r$ as expected.

The radiated power could also be calculated using the field expressions (\ref{eq:EthI}), (\ref{eq:HphI}) interior to the sphere $r=L$. Utilizing (\ref{eq:chimp}), (\ref{eq:roots1}), and (\ref{eq:Inumu1}) gives
\begin{equation}
\frac{2\pi P_{\rm rad}^-}{\eta\abs{I_0}^2} =:P_o^-= \frac{K\pi}{\eta}\Re(KY_t) = \sum\limits_{\ell =1,3}^\infty\Re(c_\ell)P_\ell(\cos\theta_0), \label{eq:PoI}
\end{equation}
where the superscript $^-$ indicates that the fields in the region $r<L$ have been used.  The operator $\Re(\cdot)$ denotes `real part' of a quantity. Evidently, $P_o^-$ is also independent of the radius $r$. It is easy to verify that the quantities on the RHSs of (\ref{eq:PoII}) and (\ref{eq:PoI}) are identical to each other by following these sequential steps:
\begin{enumerate}[(i)]
\item Start from equation (\ref{eq:cmeq}), multiply both sides of it by $-jc_m$ and sum over all $m=1,3,\ldots\infty$,
\item Take the real part of the resulting summation,
\item Utilize the symmetry relation $g_{m\ell} = g_{\ell m}$ and the Wronskian $W_H$ to arrive at the end result.
\end{enumerate}
Consequently, we set $P_{\rm rad}^+ = P_{\rm rad}^-=:P_{\rm rad}$ and $P_o^+ = P_o^- =:P_o$. Irrespective of which fields are used in the computation of radiated power, it is seen that finding the mode coefficients $c_\ell$ in the exterior region is central. 

\begin{figure}[htb]
\centerline{\scalebox{0.45}{\includegraphics{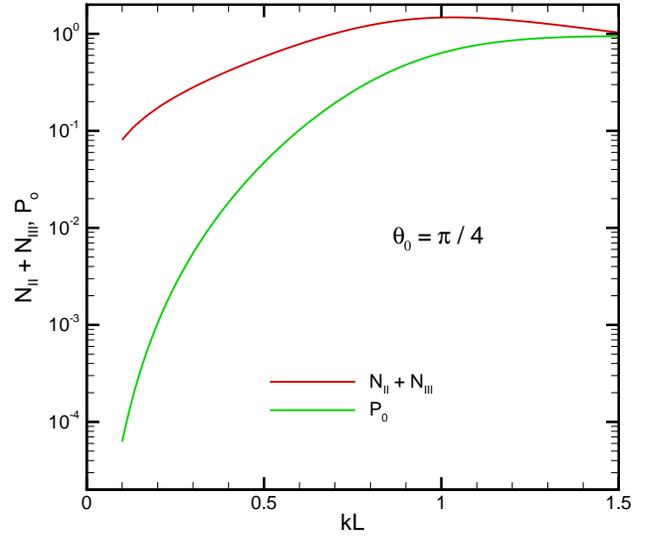}}}
\begin{center}
\caption{Normalized evanescent energy stored and normalized power radiated by the biconical antenna for $\theta_0 = \pi/4$.}
\label{fg:Pstorad}
\end{center}
\end{figure}

Fig.~\ref{fg:Pstorad} shows the variation of the normalized radiated power $P_o$ with the electric length $kL$ for a cone half-angle $\theta_0 = \pi/4$. The radiated power is very low for small electrical lengths and varies by four orders of magnitude as the length is varied over $kL\in (0.1,1.5)$.  

It is straightforward to evaluate the directivity, $D$, of the biconical antenna from expressions of the far-zone field given in (\ref{eq:Biconfar}) and the radiated power given in (\ref{eq:PoII}). The result is
\begin{equation}
D = \frac{\textstyle
\max\limits_{0<\theta<\pi}\abs{\sum\limits_{\ell=1,3}^\infty \frac{j^\ell c_{\ell}}{\widehat{H}_{\ell}^{(2)}(kL)}\frac{d P_{\ell}(\cos\theta)}{d\theta}}^2}
{\sum\limits_{\ell=1,3}^\infty\frac{\ell(\ell+1)}{(2\ell+1)}\frac{\abs{c_\ell}^2}{\abs{\widehat{H}^{(2)}_\ell(kL)}^2}
}.\label{eq:BiconDir}
\end{equation}   

\begin{figure}[htb]
\centerline{\scalebox{0.45}{\includegraphics{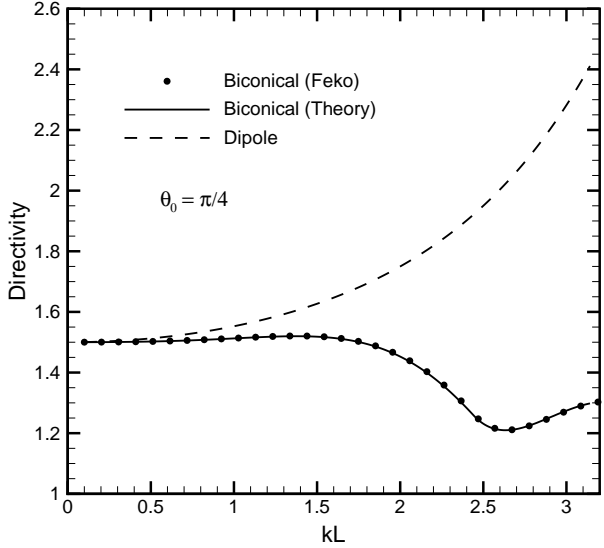}}}
\begin{center}
\caption{Directivity of a biconical antenna for $\theta_0 = \pi/4$ versus that of a cylindrical dipole having the same length.}
\label{fg:BiconDir}
\end{center}
\end{figure}

Fig.~\ref{fg:BiconDir} shows a plot of the directivity of the biconical antenna as a function of the electrical length for a fixed half-angle of $\theta_0 = \pi/4$. Simulation results obtained with Feko\footnote{Feko simulation results used 4,232 surface triangles to represent the geometry, resulting in 6,400 complex-valued unknowns. Simulation results were generated in 2.22 secs on an 8-core machine and incurred 70.5 MBs of memory \cite{CJ2021}.} are shown for comparison and are seen to agree very well with the theoretical results. Also shown for comparison is the directivity of a filamentary cylindrical dipole having the same overall length. Both antennas have the same directivity of $D=1.5$ for very small electrical lengths. However, the beam begins to split at broadside for lengths in the range  $1/4\lessapprox L/\lambda\lessapprox 1/2$ in the case of a biconical antenna, thus resulting in reduced directivity when compared to a cylindrical dipole. Nonetheless, the biconical antenna is seen to have much larger directivity bandwidth than a filamentary dipole for radii up to $L/\lambda \approx 1/4$. 

\subsection{Stored Energy and Quality Factor}
Quality factor is an important metric in the understanding of the finite bandwidth behavior of an antenna. The generally accepted definition of the quality factor, $Q$, of a lossless antenna operating in a sinusoidal steady state is \cite{chu1948}, \cite{Harr1960}
\begin{equation}
Q = \frac{2\omega}{P_{\rm rad}}\max\left[W_e,W_m\right],\label{eq:Qdef1}
\end{equation}
where $W_e\ (W_m)$ is the stored electric (magnetic) energy in the vicinity of the antenna. Because the total energy stored {\em exterior} to an antenna is infinite, one uses the evanescent energy part only in the above definition, obtained by subtracting the energy flux density associated with the radiation fields from the total energy flux density \cite{Collin}, \cite{fante}. Furthermore, the antenna is often assumed ideal and any energy stored inside the circumscribing sphere is ignored. The resulting $Q$ then sets a lower bound on the $Q$ achievable by the antenna; any energy stored within the circumscribing sphere could only increase this ideal $Q$. The evanescent electric and magnetic energies, $\widetilde{W}_e^{\rm ext}$, $\widetilde{W}_m^{\rm ext}$, respectively, outside a circumscribing sphere of radius $a$ (= $L$ in the case of the biconical antenna) are defined as \cite{Collin}, \cite{fante}
\begin{eqnarray}
\widetilde{W}_{e}^{\rm ext} &=& \frac{\epsilon_0}{4}\int\limits_{r=a}^\infty\left[r^2 \oiint\limits_{\Omega}{\bf E}\cdot{\bf E}^*d\Omega - 2\eta P_{\rm rad}\right]dr\label{eq:We_ev}\\
\widetilde{W}_{m}^{\rm ext} &=& \frac{\mu_0}{4}\int\limits_{r=a}^\infty\left[r^2 \oiint\limits_{\Omega}{\bf H}\cdot{\bf H}^*d\Omega - \frac{2}{\eta} P_{\rm rad}\right]dr\label{eq:Wm_ev}
\end{eqnarray}
Note that the same radiation flux density is extracted out from the electric and magnetic energy densities. Most works including \cite{Collin}, \cite{fante}, \cite{mclean}, \cite{Yagh2005}, \cite{gustafsson} then compute the antenna $Q$ by considering only the external evanescent energies and using the expression  
\begin{equation}
Q = \frac{2\omega}{P_{\rm rad}}\max\left[\widetilde{W}_e^{\rm ext},\widetilde{W}^{\rm ext}_m\right]=:Q^{\rm ext},\label{eq:Qdefext}
\end{equation}
For the biconical antenna expressions for fields are also available within the circumscribing sphere, see (\ref{eq:ErI})-(\ref{eq:HphI}), (\ref{eq:atob}). Therefore the structure permits one to assess the effects of internal stored energy on the computation of $Q$. While the {\em internal} stored energy is finite, one may still want to extract out the radiation flux density from the total internal flux density so as to set it on par with the exterior quantities. There is no unique way to define the evanescent {\em internal} energy. Furthermore, the difference between the internal electric and magnetic energies is not a definite quantity even if the internal modes are purely TM or purely TE\footnote{This is in contrast to the external quantities which obey certain definiteness relation, see equation (\ref{eq:We_Wm2}) and the discussion following it.}. Here we define the  total internal evanescent electromagnetic energy, $W_{em}^{\rm int}$ in the field as 
\begin{eqnarray}
W_{em}^{\rm int} &=& \frac{\epsilon_0}{4}\int\limits_{r=0}^a\Bigg[r^2 \oiint\limits_{\Omega}\left({\bf E}\cdot{\bf E}^*+\eta^2{\bf H}\cdot{\bf H}^*\right)d\Omega\nonumber\\
&& - 4\eta P_{\rm rad}\Bigg]dr>0,\label{eq:Psto}
\end{eqnarray}
which is based on the {\em total} electromagnetic energy stored inside the circumscribing sphere and one which is consistent with the definition comprising the sum of (\ref{eq:We_ev}) and (\ref{eq:Wm_ev}). A $Q$-factor based on the total stored energy is then
\begin{eqnarray}
Q^{\rm tot} &=& \frac{\omega}{P_{\rm rad}}\left(2\max\left[\widetilde{W}_e^{\rm ext},\widetilde{W}^{\rm ext}_m\right]+ W_{em}^{\rm int}\right)\nonumber\\
&=&Q^{\rm ext}+\frac{\omega W_{em}^{\rm int}}{P_{\rm rad}}.\label{eq:Qdeftot}
\end{eqnarray}
We will also compute the quality factor of the biconical antenna based on the equivalent circuit definition \cite{Yagh2005}:
\begin{equation}
Q^{\rm ckt} \approx \frac{\omega}{Z_{\rm in}+Z_{\rm in}^*}\abs{\frac{dZ_{\rm in}}{d\omega}} = \frac{kL}{2KG_{\rm in}}\abs{\frac{d(KY_{\rm in})}{d(kL)}},\label{eq:Qckt}
\end{equation}
where $Z_{\rm in} = Y_{\rm in}^{-1}$ is the input impedance of the antenna. 

Applying power conservation to the volume exterior to the sphere $r=L$ and taking into account the fact that the imaginary part of the complex power flow (\ref{eq:Pcflow}) at infinity is zero, we arrive at
\begin{eqnarray}
2\omega (W_e^{\rm ext}-W_m^{\rm ext}) &=& 2\omega (\widetilde{W}_e^{\rm ext}-\widetilde{W}_m^{\rm ext})\nonumber\\
&=&\frac{\eta\abs{I_0}^2}{2\pi}\sum\limits_{\ell=1,3}^\infty\Biggl\{\frac{\abs{c_\ell}^2\ell(\ell+1)}{(2\ell+1)}\times \nonumber\\
&&\times\ \Re\left[-\frac{\widehat{H}_\ell^{(2)\prime}(kL)}{\widehat{H}_\ell^{(2)}(kL)}\right]\Biggr\}\label{eq:We_Wm}\\
&=:&\frac{\eta\abs{I_0}^2}{\pi}N_{\rm III}>0\label{eq:We_Wm2}\end{eqnarray}
where $W_e^{\rm ext}, W_m^{\rm ext}$ are, respectively, the time-averaged electric and magnetic energies stored in the volume $r>L$. The definiteness relation in (\ref{eq:We_Wm2}) follows from the inequality \cite[C.146]{RJ2020} relating to spherical Hankel functions. Hence the stored electric energy exceeds the stored magnetic energy in the volume exterior to the circumscribing sphere $r=L$ in the case of a symmetrical biconical antenna. This remains true irrespective of the values of $L$ and $\theta_0$ of the antenna. The same however cannot be said of the electric and magnetic energies stored {\em inside} the circumscribing sphere. Indeed resonance demands that excess electric energy stored in the exterior volume $r>L$ must be balanced by excess magnetic energy stored in the interior volume $r<L$. The frequency at which the electric and magnetic energies {\em inside} the circumscribing sphere become equal to each other will be before the onset of first resonance of the antenna (that is, prior to $kL \approx 0.8$ in Fig.~\ref{fg:BCA_Admin}). 

Using (\ref{eq:ErI})-(\ref{eq:HphI}) and the results (\ref{eq:Inumu2}) and (\ref{eq:Inumu1}) it is easy to see that
\begin{eqnarray}
\oiint\limits_\Omega r^2\abs{E_r}^2\,d\Omega &=& \frac{\abs{I_0}^2\eta^2\sin\theta_0}{\pi(kr)^2}\sum\limits_{n=1}^\infty\Biggl\{\frac{\abs{a_{\nu_n}}^2}{2\nu_n+1}\nonumber\\
&&\left[\frac{\widehat{J}_{\nu_n}(kr)}{\widehat{J}_{\nu_n}(kL)}\right]^2\frac{\left(\frac{\partial M_{\nu_n}}{\partial\theta_0}\right)^2}{\frac{d\nu_n}{d\theta_0}}\Biggr\},\nonumber\end{eqnarray}
\begin{eqnarray}
\oiint\limits_\Omega r^2\abs{E_\theta}^2\,d\Omega &=& \frac{\abs{I_0}^2\eta^2\sin\theta_0}{\pi}\sum\limits_{n=1}^\infty\Biggl\{\frac{\abs{a_{\nu_n}}^2}{\nu_n(\nu_n+1)(2\nu_n+1)}\nonumber\\
&&\hspace{-0.2in}\left[\frac{\widehat{J}^\prime_{\nu_n}(kr)}{\widehat{J}_{\nu_n}(kL)}\right]^2
\frac{\left(\frac{\partial M_{\nu_n}}{\partial\theta_0}\right)^2}{\frac{d\nu_n}{d\theta_0}}\Biggr\}
+\frac{\abs{I_0}^2\eta K}{2}\bigg[1+\nonumber\\
&&\hspace{-0.2in}K^2\abs{Y_t}^2+(1-K^2\abs{Y_t}^2)\cos 2k(r-L)\nonumber\\
&&+\ 2K\Im(Y_t)\sin 2k(r-L)\bigg],\nonumber
\end{eqnarray}
\begin{eqnarray}
\oiint\limits_\Omega r^2\abs{\eta H_\theta}^2\,d\Omega &=& \frac{\abs{I_0}^2\eta^2\sin\theta_0}{\pi}\sum\limits_{n=1}^\infty\Biggl\{\frac{\abs{a_{\nu_n}}^2}{\nu_n(\nu_n+1)(2\nu_n+1)}\nonumber\\
&&\hspace{-0.2in}\left[\frac{\widehat{J}_{\nu_n}(kr)}{\widehat{J}_{\nu_n}(kL)}\right]^2
\frac{\left(\frac{\partial M_{\nu_n}}{\partial\theta_0}\right)^2}{\frac{d\nu_n}{d\theta_0}}\Biggr\}
+\frac{\abs{I_0}^2\eta K}{2}\bigg[1+\nonumber\\
&&\hspace{-0.2in}K^2\abs{Y_t}^2-(1-K^2\abs{Y_t}^2)\cos 2k(r-L)\nonumber\\
&&\hspace{-0.2in}-\ 2K\Im(Y_t)\sin 2k(r-L)\bigg].\nonumber
\end{eqnarray}
Adding these three quantities and extracting out the radiated power per (\ref{eq:Psto}), one gets after some simplifications the following integral for region-I
\begin{eqnarray}
\int\limits_{\zeta = 0}^{kL}d\zeta\oiint\limits_{\Omega}\left(r^2[{\bf E}\cdot{\bf E}^*+\eta^2{\bf H}\cdot{\bf H}^*] - 4\eta P_{\rm rad}\right)d\Omega=&&\nonumber\\
\frac{\abs{I_0}^2\eta^2\sin\theta_0}{\pi}\sum\limits_{n=1}^\infty\Biggl\{\frac{\abs{a_{\nu_n}}^2}{(2\nu_n+1)}\frac{\left(\frac{\partial M_{\nu_n}}{\partial\theta_0}\right)^2}{\frac{d\nu_n}{d\theta_0}}\frac{1}{\widehat{J}_{\nu_n}^2(kL)}\times&&\nonumber\\
\int\limits_0^{kL}\left[\frac{\widehat{J}_{\nu_n}^{\,2}(\zeta)}{\zeta^2} + \frac{\widehat{J}^{\,2}_{\nu_n}(\zeta)}{\nu_n(\nu_n+1)}+\frac{[\widehat{J}_{\nu_n}^{\prime}(\zeta)]^2}{\nu_n(\nu_n+1)}\right]\,d\zeta\Biggr\}&&\nonumber\\
+\hspace{1ex} \abs{I_0}^2\eta K\abs{1-KY_t}^2kL\hspace{0.5in}&&\\
>0,\hspace{2.5in}\nonumber
\end{eqnarray}
where the substitution $\zeta = kr$ was used. 
The integral involving Bessel functions can be evaluated in a closed form using integration by parts, followed by the identity \cite[C.139]{RJ2020} to finally result in the following expression for the normalized evanescent energy stored in region-I:
\begin{eqnarray}
\frac{2\pi \omega W_{em}^{\rm int}}{\eta \abs{I_0}^2} =: N_{\rm I}&=&\frac{\sin\theta_0}{2}\sum\limits_{n=1}^\infty\Biggl\{\frac{\abs{a_{\nu_n}}^2}{(2\nu_n+1)}\frac{\left(\frac{\partial M_{\nu_n}}{\partial\theta_0}\right)^2}{\frac{d\nu_n}{d\theta_0}}\nonumber\\
&&\left[\frac{kL}{\nu_n(\nu_n+1)}\left(1+\frac{\widehat{J}^{\prime\, 2}_{\nu_n}(kL)}{\widehat{J}_{\nu_n}^{\,2}(kL)}\right)-\frac{1}{kL}\right]\Biggr\}\nonumber\\
&& +\hspace{1ex}\frac{\pi K}{2\eta }kL\abs{1-KY_t}^2.\label{eq:PstoI}
\end{eqnarray}
For $kL\to 0$, asymptotic forms of Bessel functions can be used to result in the asymptotic expression 
\begin{equation}
N_{\rm I}\sim\frac{\sin\theta_0}{2 kL}\sum\limits_{n=1}^\infty \frac{\abs{a_{\nu_n}}^2}{\nu_n(2\nu_n+1)}\frac{\left(\frac{\partial M_{\nu_n}}{\partial\theta_0}\right)^2}{\frac{d\nu_n}{d\theta_0}},\ kL\to 0. \label{eq:PstoIasmp}
\end{equation}
Note that the mode coefficients $a_{\nu_n}$ are themselves dependent on $kL$ through the coefficients $c_\ell$ as is evident from (\ref{eq:atob}) and (\ref{eq:cmeq}). Note also that one can avoid the explicit calculation of the quantity $\partial M_{\nu_n}/\partial\theta_0$ that is needed in (\ref{eq:PstoI}). It is indeed clear from (\ref{eq:atob}) and (\ref{eq:Inumu2}) that $\sin\theta_0\abs{a_{\nu_n}}^2(\partial M_{\nu_n}/\partial\theta_0)^2/[d\nu_n/d\theta_0] = \nu_n(\nu_n+1)(2\nu_n+1)\sum_\ell\sum_m c_\ell c^*_mm(m+1)\ell(\ell+1)T_{\ell,\nu_n}(\theta_0)T_{m,\nu_n}(\theta_0)/I_{\nu_n,\nu_n}(\theta_0)$ and that the ratio $T_{\ell,\nu_n}(\theta_0)T_{m,\nu_n}(\theta_0)/I_{\nu_n,\nu_n}(\theta_0)$ is independent of $\partial M_{\nu_n}/\partial\theta_0$. The evanescent energy in region-I, a positive definite quantity, is completely determined once the mode coefficients $c_\ell$ in the exterior region have been found. 

In a similar fashion using the expressions (\ref{eq:ErII})-(\ref{eq:HphII}) for the fields in region-II and the identities (\ref{eq:PlPn}) and (\ref{eq:dPldPn}) it can be shown that with $\zeta = kr$ 

\begin{eqnarray}
r^2\oiint\limits_{\Omega}\left[{\bf E}\cdot{\bf E}^*+\eta^2{\bf H}\cdot{\bf H}^*\right]d\Omega=\frac{\abs{I_0}^2\eta^2}{\pi}&&\nonumber\\
\sum\limits_{\ell=1,3}^\infty\Biggl\{\frac{\ell(\ell+1)\abs{c_{\ell}}^2}{(2\ell+1)\abs{\widehat{H}_{\ell}^{(2)}(kL)}^2}\times&&\nonumber\\
\bigg[\ell(\ell+1)\frac{\abs{\widehat{H}_{\ell}^{(2)}(\zeta)}^2}{\zeta^2} + \abs{\widehat{H}^{(2)\,\prime}_{\ell}(\zeta)}^2+\abs{\widehat{H}^{(2)}_{\ell}(\zeta)}^2\bigg]\Biggr\}.&&\nonumber
\end{eqnarray}

From this and the identities \cite[C.139, C.154]{RJ2020}, the expression for the total evanescent energy stored in region-II is obtained as 
\begin{eqnarray}
\frac{2\pi\omega (\widetilde{W}_e^{\rm ext}+\widetilde{W}_m^{\rm ext})}{\eta \abs{I_0}^2}&=&
\frac{kL}{2}\sum\limits_{\ell=1,3}^\infty\Biggl\{\frac{\ell(\ell+1)\abs{c_{\ell}}^2}
{(2\ell+1)\abs{\widehat{H}_{\ell}^{(2)}(kL)}^2}
\nonumber\\
&&\bigg[\left(\frac{\ell(\ell+1)}{(kL)^2}-1\right)\abs{\widehat{H}_{\ell}^{(2)}(kL)}^2 + \nonumber\\
 &&2-\abs{\widehat{H}^{(2)\,\prime}_{\ell}(kL)}^2\bigg]\Biggr\}.\label{eq:PstoII}\\
 &=:&N_{\rm II}
\end{eqnarray}
Note that $N_{\rm II}>0$ for any $kL$ in view of the inequality \cite[C.148]{RJ2020}. The sum of $N_{\rm II}$ and $N_{\rm III}$, which represents the evanescent electric energy stored in the exterior volume, is thereby a positive definite quantity for any $kL$ and $\theta_0$. 

In terms of the normalized quantities $N_{\rm I}, N_{\rm II}, N_{\rm III}, P_o$, the various quality-factors of the antenna are 
\begin{eqnarray}
Q^{\rm ext} &=& \frac{(N_{\rm II}+N_{\rm III})}{P_o},\label{eq:QBCA2}\\
Q^{\rm tot}&=&Q^{\rm ext}+\frac{N_{\rm I}}{P_o}\label{eq:QBCA3}\end{eqnarray}

The sum of the quantities $N_{\rm III}$ and $N_{\rm II}$ is also plotted in Fig.~\ref{fg:Pstorad} for $\theta_0 = \pi/4$. It is seen that the dynamic range in the variation of the sum is far less than that of the radiated power. 

\begin{figure}[htb]
\centerline{\scalebox{0.45}{\includegraphics{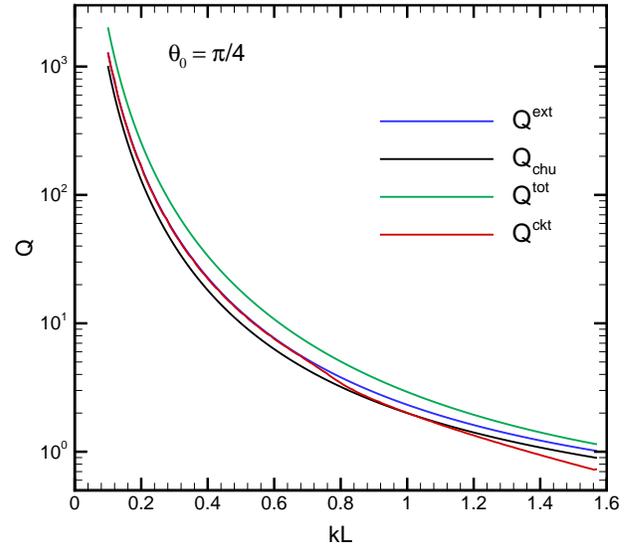}}}
\begin{center}
\caption{Quality factor of a biconical antenna for $\theta_0 = \pi/4$ by various formulations. Comparison is shown with Chu's limit $Q_{\rm chu} = (kL)^{-1}[1+(kL)^{-2}]$.}
\label{fg:Qbca}
\end{center}
\end{figure}

Fig.~\ref{fg:Qbca} shows the $Q$-factor for the biconical antenna with $\theta_0 = \pi/4$. Comparison is shown with the Chu's limit of \cite{mclean}
\begin{equation}
Q_{\rm chu} = (kL)^{-1}[1+ (kL)^{-2}],\label{eq:QChu} 
\end{equation}
which is the minimum value achievable by any lossless antenna circumscribed within a sphere of radius $L$. In general, a high $Q$ and a high slope at smaller antenna lengths is a consequence of the relatively low radiated power as evident from Figure~\ref{fg:Pstorad}. It is seen that the values of $Q$ calculated by external evanescent energy (\ref{eq:QBCA2}) and that by the circuit model (\ref{eq:Qckt}) follow each other quite closely until $kL \lessapprox 0.7$, while that calculated by the total evanescent energy (\ref{eq:QBCA3}) exceeds both of them. Finally, the three values are higher than the minimum value predicted by (\ref{eq:QChu}) in the range $0.1\le kL\lessapprox 0.7$. For instance, at $kL=0.5$,\ $Q_{\rm chu} = 10$,\ $Q^{\rm ext} = 12.2$,\ $Q^{\rm tot} = 17.6$, $Q^{\rm ckt} = 12$. In contrast, at $kL = 0.1$, we have $Q^{\rm ext}/Q_{\rm chu} = 1.278,\ Q^{\rm ckt}/Q_{\rm chu} = 1.268,\ Q^{\rm tot}/Q_{\rm chu} = 2$. The $Q$-factor calculated by the circuit model, however, breaks down when $kL>0.7$ and even falls below the Chu's limit of (\ref{eq:QChu}) for $kL\gtrapprox 1$. The consolation is that this breakdown happens at lengths where the significance of the $Q$-factor diminishes.

\begin{figure}[htb]
\centerline{\scalebox{0.45}{\includegraphics{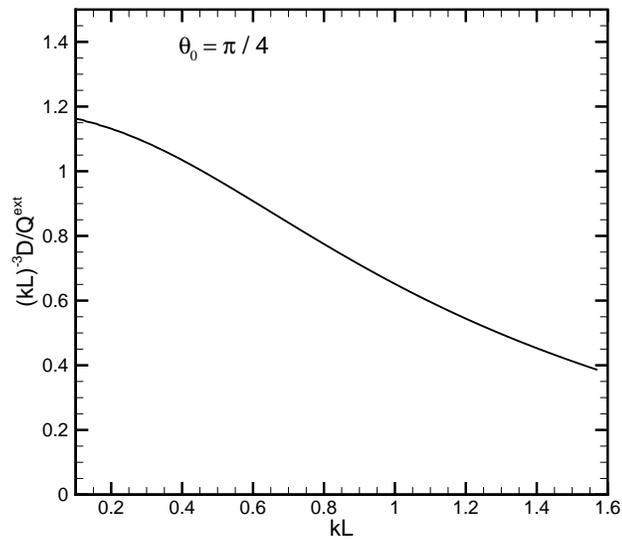}}}
\begin{center}
\caption{The figure of merit $(kL)^{-3}D/Q^{\rm ext}$ of a biconical antenna versus electrical length.}
\label{fg:BCA_DbyQ}
\end{center}
\end{figure}

Finally Fig.\ref{fg:BCA_DbyQ} shows a plot of the antenna figure of merit $F = (kL)^{-3}D/Q^{\rm ext}$ versus $kL$ for a biconical antenna. For omnidirectional antennas, an upper  bound to this figure of merit is \cite{chu1948}, \cite{gustafsson3} $F\le 1.5$. It is seen that the figure of merit of an electrically {\em small} biconical antenna differs from the ideal upper bound by a factor of $0.775\approx 0.78$. The figure of merit deteriorates more relative to the ideal number at larger lengths.  

\section{Conclusion}
Complete expressions have been provided for the analysis of a biconical antenna of arbitrary cone angle and length. Numerical data have been generated for various quantities for a cone angle of $\theta_0 = \pi/4$. A wide-angle biconical antenna has many desirable features and the following steps summarize the key findings of the study:

\begin{enumerate}[(i)]
\item Equations (\ref{eq:Rootsasymp1}) and (\ref{eq:Rootsasymp2}) provide highly accurate analytical expressions for the roots $\nu_n(\theta_0)$ and the derivative $d\nu_n/d\theta_0$ for an arbitrary cone angle. This has been established by favorable comparison with numerically generated ones.
\item The mode coefficients $c_\ell$ are determined from the matrix equation (\ref{eq:BCA_Coeff}) and (\ref{eq:vecp}) with the matrix entries of $\cal G$ filled from (\ref{eq:CalG}) and (\ref{eq:Newgml}). The latter provides a convergent means for evaluating the matrix entries. 
\item The input admittance $Y_{\rm in}$ of the antenna is determined from (\ref{eq:BiconInAdm}), (\ref{eq:Gamin}) and (\ref{eq:BiconAdm}). The admittance data has been validated via commercial software package WIPL-D. Numerical results for the admittance variation with electrical length for $\theta_0=\pi/4$ indicate that the antenna is broadband for radii $L/\lambda\gtrapprox 1/4$. Furthermore, susceptance data in Fig.~\ref{fg:BCA_Admin} clearly demonstrate that Foster's reactance theorem remains invalid even for perfectly conducting antennas. 
\item The directivity, $D$,  of the antenna is determined from (\ref{eq:BiconDir}). Results for $\theta_0=\pi/4$ indicate that the directivity of the antenna is a slowly varying function of its electrical length and departs significantly from that of a filamentary cylindrical dipole having the same electrical length. The directivity has been validated via commercial package FEKO. 
\item The quality factor of the biconical antenna can be computed by one of (\ref{eq:QBCA2}), (\ref{eq:QBCA3}), or (\ref{eq:Qckt}). Results for $\theta_0=\pi/4$ indicate that if the energy within the circumscribing sphere is ignored, the $Q$-factor of the antenna differs from the lower limit established by Chu roughly by a factor of 1.3. In addition, the simple formula (\ref{eq:Qckt}) based on the frequency derivative of input admittance generates results that closely match those from formulation (\ref{eq:QBCA2}). If internal energy is included in the computation of $Q$ as in formulation (\ref{eq:QBCA3}), the quality factor differs from the lower limit roughly by a factor of 2. This is by far the most realistic case for a biconical antenna.
\item For a biconical antenna with $\theta_0=\pi/4$, the figure of merit $F = (kL)^{-3}D/Q^{\rm ext}\le 1.1624$ over the range $kL\in(0.1,\pi/2)$. This is in contrast to an {\em ideal} omnidirectional antenna which satisfies a higher upper bound of $F\le 1.5$.
\end{enumerate}

% or


%\appendices


%\section{}

\section*{Acknowledgment}
The author would like to thank Dr. B. Mrdakovic and Dr. C. J. Reddy for sharing the WIPL-D and Feko results shown in Figs.~\ref{fg:BCA_Admin} and \ref{fg:BiconDir}, respectively.  

\ifCLASSOPTIONcaptionsoff
  \newpage
\fi



%\bibliographystyle{IEEEtran}
\bibliographystyle{IEEEtran}
\bibliography{BiconRef.bib}


%\begin{IEEEbiography}
%[{\includegraphics[width=1in,height=1.5in,clip,keepaspectratio]{IMG_0013.jpg}
%}]{Ramakrishna Janaswamy is a professor in the electrical and computer engineering department at the University of Massachusetts at Amherst. He served as an Associate Editor for several professional journals and currently an active member of the IEEE Antennas \& Propagation Standards Committee. His research interests are in the areas of analytical and computational electromagnetics, mathematical physics, and system theory.}
%\end{IEEEbiography}


%\vfill


\end{document}